# Design and Fabrication of Industrially Scalable low cost Liquid Impregnated Surfaces with extreme hydratephobic properties

Amit K Nayse, Abhishek Mund, Arindam Das*

School of Mechanical Sciences, Indian Institute of Technology (IIT) Goa, GEC Campus, Farmagudi, Ponda, Goa, 403401, India

*Email: arindam@iitgoa.ac.in

## Abstract

The design and fabrication of extremely hydrate phobic Liquid Impregnated Surfaces (sometimes abbreviated as LIS) based on industrial material Aluminium Al6061 and industrially scalable low-cost method were carried out. A simple hydrochloric acid-based etching method and boiling water treatment were used to generate micro and nanoscale nanopetal roughness features respectably. A theoretical analysis was performed to find out the relationship between the interfacial interactions and surface roughness features to predict the stability of lubricant oil of LIS under water and oil environment. LIS with appropriate surface chemistry and textures were fabricated to experimentally validate the theoretical analysis on lubricant stability. Subsequent experimental measurements of hydrate adhesion were performed on LIS with stable lubricant layers, using a custom-made experimental setup and cantilever-based method. Cyclopentane hydrate, mimics gas hydrate forming mechanism at atmospheric pressure, and is used for the hydrate adhesion measurements. Extreme reduction of hydrate adhesion with more than four orders of magnitude was observed on LIS compared to control smooth aluminium surfaces.

1. Introduction

Clathrate hydrates are solid structures resembling ice that develop when hydrophobic guest molecules are present in a water environment.[1] Their discovery dates back to 1810, credited to Sir Humphry Davy, and since then, particularly gas hydrates have been extensively explored.[2] These hydrates have been identified in various natural[3] and artificial settings. These hydrates are prevalent within oil and gas infrastructure, such as pipes and valves.[4] The conditions in these systems, moderately high pressure of around 5MPa and relatively low temperature below 10°C, are conducive to the formation of hydrate particles. Once initiated, hydrate phase rapidly grows from small particles to particle agglomerates, eventually leading to the formation of solid hydrate deposits or plugs within pipes.[1] The emergence of hydrate particles within these components gives rise to flow assurance challenges and can even result in severe accidents.[5] Addressing these issues necessitates the use of methods like anti-agglomerates, kinetic inhibitors, thermal heating systems, and mechanical depressurization systems. Regrettably, these techniques have notable environmental impacts and demand substantial energy consumption.[6] For example, methanol, an inhibitor of hydrate, is used in large quantities for effective results. It not only increases material cost but also necessitates a complicated and involved separation process to get rid of these chemicals from the oil and gas phase.[7] Hence, some environmentally friendly and energy-independent passive approach is required to effectively tackle these problems.[8] Engineered surfaces such as superhydrophobic, superoleophobic, Slippery Liquid Infused Porous Surfaces (SLIPs), and LIS have shown tremendous potential in reducing the ice adhesion[9], reducing the corrosion[10], reducing drag forces[11], decreasing scale formation[12], facilitating robust condensation[13], and increasing condensation heat transfer.[14] Due to the modified solid liquid or solid-solid contact area and low inherent adhesion forces, these surfaces possess the above-mention properties. These surfaces are expected to perform similarly in reducing the hydrate adhesion with solid pipe walls in a passive and energy-independent way.[15] In recent years, several research works as shown superior performance of superhydrophobic surfaces[16]. Under-oil superhydrophobic surfaces[17] in reducing hydrate adhesion. LIS being one of the high-performing engineered surfaces, still not being investigated to achieve reduced hydrate adhesion.[18] Properly designed LIS have shown to reduce ice adhesion[9], decreased scaling[16], and reduced drag forces.[11] Compared to the superhydrophobic and SLIPs, LIS expected to perform better due to the presence of lubricant layer which is thermodynamically stable as well as having higher density and viscosity compared to the air phase found in superhydrophobic surfaces.[19] Hence it is worth evaluating these surfaces as potential solutions for hydrate problems. A recent study from our group has shown excellent performance of LIS in reducing the hydrate adhesion.[20] However, the materials and methods used to make LIS were costly and not scalable.[21] Moreover, the presence of nano feature on top of micro features was must to achieve extremely low hydrate adhesion in those LIS. Due to the limitation of the selected surface chemistry, reported LIS based on micro-only texture showed reduction of hydrate adhesion force only by one order of magnitude compared to the smooth surface.[22] Due to the fragile nature of nano features it is desirable to have LIS based on micro-only features.[23] Hence, the selection of the appropriate surface chemistry is expected to reduce the hydrate adhesion on micro-only LIS by increasing the critical post spacing required for lubricant stability.[24]

In this current work, for the first time, an industrially scalable method was reported to fabricate aluminium 6061 alloy-based LIS with an exceptionally low hydrate adhesion force of less than 0.04 mN for a 0.5 mm diameter hydrate particle. To measure the hydrate adhesion, cyclopentane hydrate was selected as it mimics the gas hydrate forming mechanism at the hydrophobic hydrophilic interface under atmospheric pressure. A novel nanoscale conforming coating based on the Teflon solution was dip coated with the optimize speed for fabrication of LIS. A low-cost acid etching and boiling water treatment were employed to generate micro-only and micro-nano petals hierarchical features. Variations of acid concentration and etching time showed the presence of roughness features of different length scales. Based on the theoretical analysis of lubricant stability, a surface with micro features and another with a micro-nano petal hierarchical structure were selected for subsequent Teflon coating and LIS fabrication. Both of these surfaces show exceptionally low hydrate adhesion force below 0.04 mN.

## 2. Theoretical analysis

LIS are smart textured surfaces where a liquid remains inside the textures in a thermodynamic equilibrium state under a given fluid (immiscible with impregnated fluid) environment. LIS investigated so far consists of three-phases mainly consisting of probe liquid, lubricant liquid, and textured solid surface. LIS surfaces achieve repellency to probe liquid by using different oils as lubricants, thus restricting stability analysis with only two liquids and solids. However, for the hydrate repellency in an oil-water system, the choice of lubricant must be made in such a way that it should be stable beneath two environmental liquids and immiscible with them as well (oil and water). Any theoretical analysis for such a system requires an understanding of interfacial interactions involving four-phase consisting of a textured surface, lubricant liquid, and two environmental liquids. The theoretical analysis was carried out to identify the stability conditions by providing a concise understanding of how to choose a texture, lubricant, and surface chemistry based on the properties of two environmental liquids (water and oil). One critical factor determining the usability of these surfaces in real-life conditions is the robustness and stability of the impregnated fluid layer. The objective of this analysis is to find-out the criteria under which a LIS is expected to have a stable lubricant layer under both ambient liquids, which minimizes solid-to-liquid contact drastically. The thermodynamically stable interfacial configuration under distinct conditions are governed by the correlation between surface roughness and surface chemistry parameters. These correlations were established using the principle of surface energy minimization. Total surface energy is the sum of the products of interfacial areas and respective specific interfacial energies of all interfaces between different phases. The only factors affecting these two terms are geometry and surface chemistry. It is necessary to define and contrast relevant representations of geometric properties and surface chemistry in order to build the stability criterion for multiphase interfaces on LIS surfaces.

### 2.1 Geometrical Parameters

Properly defined key geometrical parameters are essential to carry out the theoretical formulation. The roughness ratio, solid fraction, and critical contact angles are essential key surface parameters selected for the current analysis. For the sake of simple analysis, surfaces with square-shaped micro posts arranged in a rectangular array with different spacings were considered, as shown in figure 1. Key roughness parameters for such surfaces are given by the following equations,

The roughness ratio; $\quad r_s = \frac{(a+b)^2 + 4ah}{(a+b)^2}$ ……………………….(1)

The solid fraction; $\quad \varphi = \frac{(a)^2}{(a+b)^2}$ …………………………….(2)

The critical contact angle; $\quad Cos(\theta_c) = \frac{1-\varphi}{r_s - \varphi}$ …………………………..(3)

Where roughness ($r_s$) is the ratio of total area and projected surface area, solid fraction ($\varphi$) is the percentage of solid contact area on projected area, ($\theta_c$) is the critical contact angle and geometry parameters width (a), edge-to-edge spacing (b), and height (h).

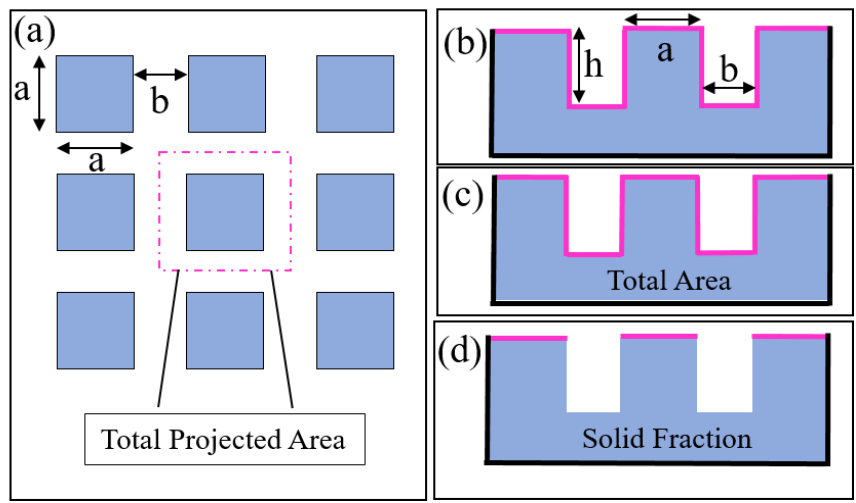

Figure 1: Micro post surfaces with key surface roughness parameters

## 2.2 Interfacial property

Specific surface energy values between liquid phases and solid surfaces are required to compute the total surface energy. These particular surface energies (interfacial forces) directly manifest intermolecular forces between multiple molecules. The equilibrium contact angles on smooth surfaces are determined by Young's equation when these interfacial surface forces balance each other at three phase contact lines, as shown in the figure 2. The Solid–Gas, Solid–Liquid, and Liquid–Gas interfacial energy denoted by $\gamma_{SG}, \gamma_{SL}$ and $\gamma_{LG}$ and $\theta_e$ is the equilibrium contact angle. This contact angle must be appropriately specified to accurately characterize the interfacial interactions between lubricant oil, environment liquids, and the solid surface.

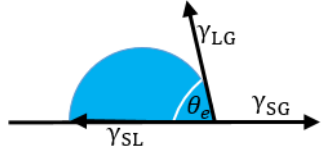

Figure 2: Water droplet on a homogeneous solid

### 2.3 Thermodynamic Stability of Lubricant oil layer

LIS surfaces have shown more non-wetting properties than superhydrophobic surfaces, such as minimal hysteresis and long-term durability.[18] The basis to their exceptional performance is the inclusion of lubricant inside the texturing, which lessens contact with solid surfaces when probing liquid is used. To find out the conditions for lubricant stability, total interfacial energies of all possible interfacial configurations involving all the phases were calculated. Figure 3 shows all such possible configurations. To find out the condition for which a particular configuration will be in the thermodynamically stable state, the total interfacial energy associated with that particular configuration was set to less than the energy associated with the other two configurations. The resulting two inequalities, when expressed in terms of roughness and surface chemistry parameters, give the necessary condition for which that specific configuration will indicate thermodynamically stable state.

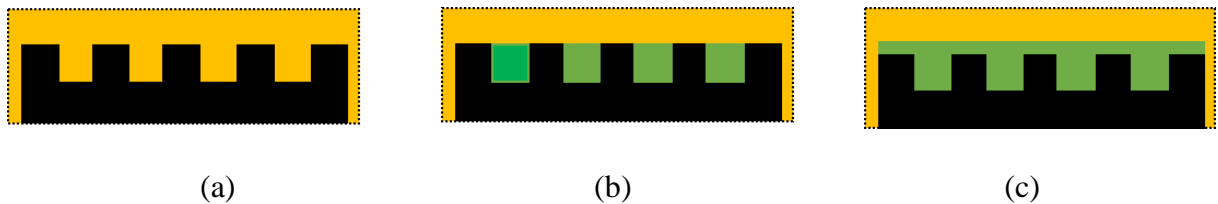

(a)  (b)  (c)

Figure 3: Configuration lubricant on the textured surface under CP environment (a) $E_1$, (b) $E_2$ and (c) $E_3$. Here green, orange, and black colours represent lubricant, cyclopentane, and solid surface, respectively.

The figure above shows three possible configurations of interface involving lubricant, environment (cyclopentane/water), and solid textured surface. Total interfacial energies associated with each configuration were given by the following equations,

$$E_1 = r_s \gamma_{SC} \quad \text{...............................................(4)}$$

$$E_2 = (1-\varphi)\gamma_{OC} + (r_s - \varphi)\gamma_{OS} + \varphi\gamma_{SC} \quad \text{.............................(5)}$$

$$E_3 = \gamma_{oc} + r_s\gamma_{os} \quad\quad\quad\quad\quad\quad\quad\quad\quad\quad\quad\quad (6)$$

$E_1$, $E_2$ and $E_3$ are total interfacial energy per unit area for 1, 2, and 3 configurations, respectively. Total surface energies for all three configurations are a function of surface energies and roughness properties. At static conditions, the lubricant oil is expected to be stable inside the texture if the total surface energy associated with either of the configurations given by figure 2. b and 2. c is minimum compared to the other two configurations. If the total energy associated with the configuration given by figure 2.a is minimum, then the lubricant will not be stable, and replaced by environmental fluid. The condition for which figure 2.b to be the thermodynamically stable configuration can be expressed in the following equation,

$$\theta_{o-s(cp)} < \cos^{-1}\left(\frac{1-\phi}{r_s-\phi}\right) \quad\quad\quad\quad\quad\quad\quad\quad (7)$$

Here $\theta_{0-s(cp)}$ is the young contact angle of lubricant oil on a smooth solid surface in a cyclopentane environment, and it is purely a function of interfacial interactions. The term at the right side of this inequality is a purely surface roughness parameter sometimes referred as critical contact angle. The figure 2.c corresponds to the condition of a fully wetted smooth surface when $\theta_{0-s(cp)} = 0$.

The stability study mentioned above, which is applied to surfaces with micron post spacings, may also be extended to surfaces with micro-nanoscale post spacings. In this work, micro-nano hierarchical structure is available on top of that micro roughness. By employing the same analogy i.e ECA to determine the stability criteria, we may expand the aforementioned study. The existence of high aspect ratios in these micro-nano hierarchical structures is responsible for the stability of oil, since they effectively lower the $\theta_{os\,(cp)}$ and increase stability. Contact angle values on these surfaces is close to zero since these micro-nano hierarchical structures have a low solid fraction. In these kinds of structures, the updated contact angle value may ensure a stable oil layer.

## 3. Results and Discussion

### 3.1 Selection of lubricant oil

Lubricant oil is the essential phase of LIS which provides unique properties such as increased liquid mobility, low contact angle hysteresis, repellency, and reduction in solid adhesion. The current application requires repellency and stability of lubricant with both water and oil phases. It becomes very challenging to find such lubricant oil, which also has to be immiscible to both water and oil phases. Fluorinated oils such as krytox 1506 and GPL 102 possess this unique property of simultaneous immiscibility with water and oil phases and are selected as lubricant oil.

### 3.2 Selection of surface chemistry

Surface chemistry and interfacial interactions between various phases are extremely important to the development and stability of LIS. Contact angles are an essential parameter because they represent surface chemistry and interfacial interactions. The most decisive element for stability analysis is the contact angles of lubricant oil krytox on smooth solid surfaces under both water and cyclopentane environments. It is evident from the theoretical analysis that lower the contact angle of lubricant oil on a smooth surface in a given liquid environment, higher the post spacings, which can keep it in thermodynamically stable condition. Our recent work reported on hydrate adhesion over LIS based fluorosilane (FS) treated silicon micropost surfaces showed only one order of reduction in hydrate adhesion. Adhesion forces were still significant due to the higher lubricant contact angles (on smooth FS surface), which allows a stable lubricant layer only on closley spaced micro post silicon surfaces. In this work, a liquid teflon-based surface coating was used to achieve further reduction on this contact angle. Experimental observation reveals a very small contact angle of krytox on teflon coated smooth surfaces under both environmental liquids. The measured Equilibrium Contact Angle (ECA) values of Krytox 1506 on smooth Teflon-coated Al surfaces under cyclopentane and water environments are $10° \pm 2°$ and $12° \pm 1°$ respectively (see figure 4). It suggests that LIS based on Teflon coated will have a thermodynamically stable lubricant (krytox) layer in both a water and cyclopentane environment. The theoretical calculations show teflon coated micropost surfaces with post spacings up to 95 microns can hold krytox under water and cyclopentane environment. Hence teflon coating was selected as surface chemistry for LIS fabrication.

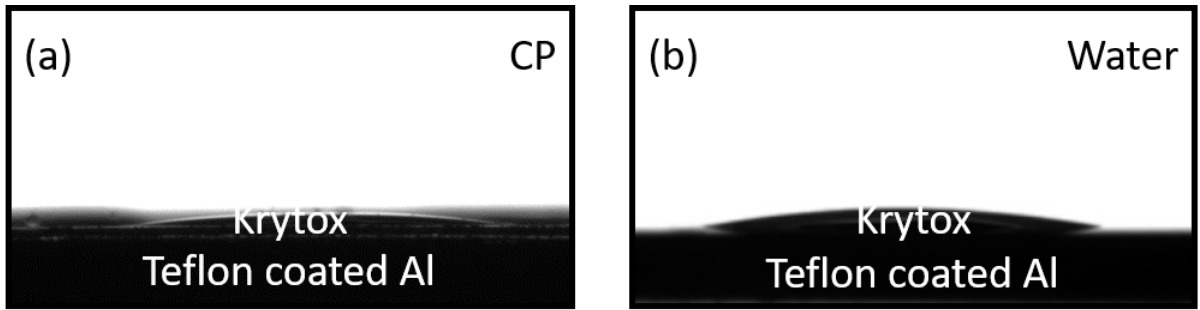

Figure 4: Krytox 1506 drop on smooth Teflon coated Al surface under (a) cyclopentane and (b) water environment

### 3.3 Surface texture characterization and selection

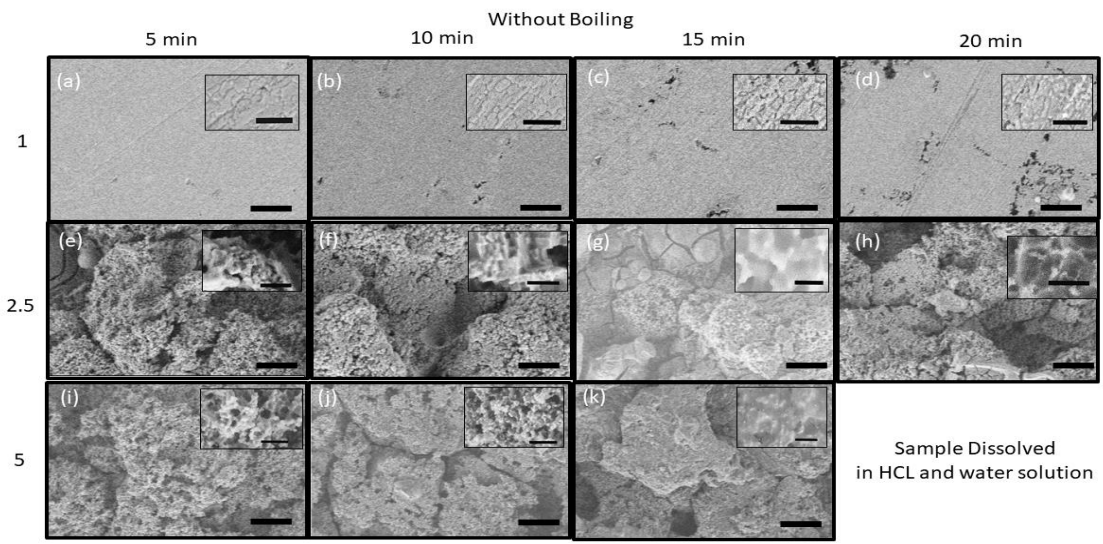

Figure 5: FE-SEM images of different acid concentrations and times. Scale bars are 5 micrometer in length and 500 nm in length (inset).

The choice of appropriate surface texture is crucial for fabricating stable LIS. The current investigation employed acid etching and boiling water treatment of aluminum 6061 alloys to generate micro or micro-nano scale hierarchical surface textures for LIS fabrication. Details of this process have been described in the materials and method section. Field Emission Scanning Electron Microscopic (FE-SEM) images were taken for surface characterization and used to identify surfaces with suitable features required for stable LIS. FE-SEM Images of Acid Etched (AE) and aluminium 6061 surfaces are shown in figure 5 for different acid concentrations and different acid etching times. Figure 5 (a) to (d) show that surfaces treated with one molar HCl solution have no significant microscale features. Only some microscale voids appeared in these samples, and their number increased slightly with an increase in acid etching time. High-resolution images of these samples showed the absence of significant nano-scale features except nano-scale cracks. The depth of these nanoscale cracks appeared to be increasing with longer acid etching time. These nanoscale cracks are most probably representing grain boundaries which are more susceptible to chemical attack by acid. FE-SEM images of samples treated with 2.5 molar HCl solutions showed significant micro-nano scale hierarchical surface morphologies (see figure 5.(e) to 5.(h)). With increased acid etching time a significant increase in microscale roughness was observed along with increased porosity, decreased width, and higher pick-to-valley distance in micro features. High-resolution FE-SEM image of the submicron range reveals the presence of nanoscale features in all four samples. Nanoscale features were more prominent on surfaces treated for 10 minutes and comprising of layered morphology as shown in the inset figure 5 (f). Further increase in acid etching time Seems to smoothen out the nanoscale features as shown in the inset images of 5 (g) and 5 (h). FE-SEM micrograph of samples treated with 5 molar HCl with 5 minutes etching time shows highly hierarchical surface features comprising of porous micro features (see figure 5 (i)). Samples with higher etching time show micro features with 2D flaky and less porous morphology, as shown in figure 5.(j) and (k). Due to the vigorous nature of the chemical reaction substrate in these two cases was reduced to a thickness of less than 0.5 mm. The reaction rate for this particular concentration was extremely vigorous, and the entire substrate of 2 mm thickness was dissolved in the acid solution within 20 minutes of reaction time. High-resolution images taken at sub-micron scale reveal the presence of spherical nano feature on samples with acid treatment time of 10 minutes or less, as shown in inset images 5.(i) and (j). Further, increase in etching time completely smoothens out nanoscale feature as shown in figure 5.(k). The sample treated for more than 5 minutes were mainly comprised of loosely attached porous micro features. The thickness of the samples was reduced to less than 0.5 mm. FE-SEM images taken at sub-micron scale reveal the presence of particulate type nano features in samples with an acid etching time 10 minutes or less. Nano features smoothen out or completely disappear when acid etching time increases beyond 10 minutes as shown in the inset image 5.(k).

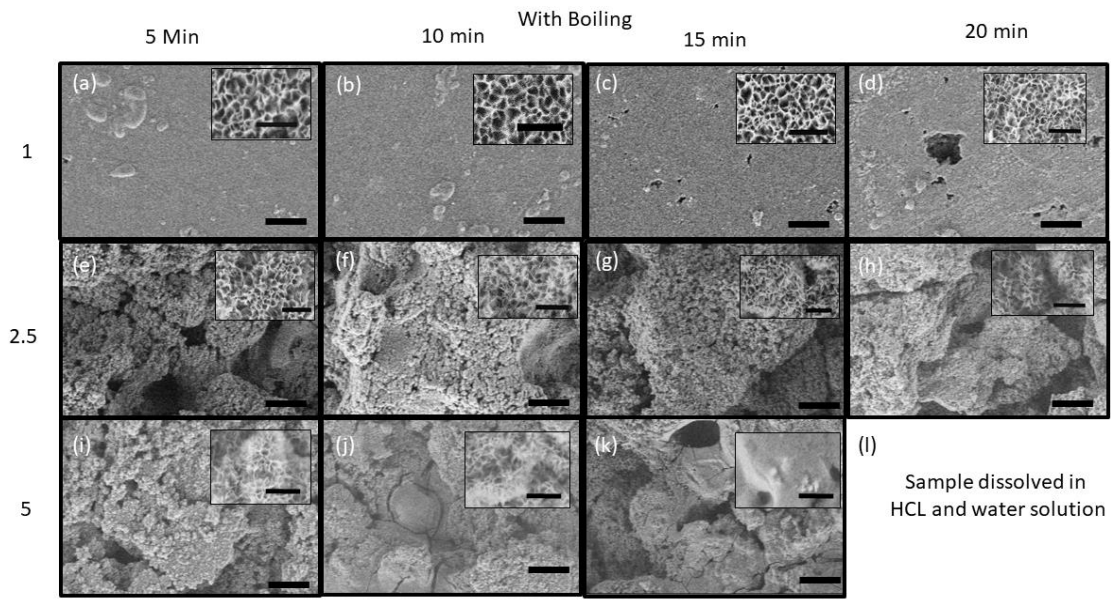

Figure 6: FESEM images of different acid concentrations and times with boiling water treatment. Scale bar are 5 micrometer and 500 nm (inset).

In a subsequent study, the effects of 20 minutes boiling water treatment on the surfaces mentioned in the previous paragraph were analysed through FE-SEM images, as shown in figure 6 above. Boiling water treatment on samples etched with one molar acid solution seems to generate micron-scale spherical surface features. The density of these micro features increased on surfaces with higher acid etching time. Microscale grooves which were present in AE samples seem to exist even after boiling water treatment. Submicron scale images of these one molar AE sample (figure 6. (a) to (d)) reveals the presence of densely packed nanosheets or nano petals morphology after boiling water treatment. The density of these nano petals morphology remain unchanged on samples with different etching times (1M HCl solution). The typical size of voids between these nano petals and the height of nano petals were around 100 nm. In contrast, the thickness of nano petals was in the range of 20 to 30 nanometres. FE-SEM images of surfaces etched with 2.5 Molar solution and subsequent boiling water treatment shows no significant effect of boiling water treatment in microscale morphology (see figure 6. (e) to (h)). High-resolution images (see inset image 6. (e) to (h)) on these samples reveal the presence of nano-petal morphology similar to what was observed in all surfaces treated with one molar acid solution followed by boiling water treatment. Acid etching time seems to have no effect on nanoscale morphology for these samples treated with 2.5 molar acid solutions and subsequent boiling water treatment. Nano petals morphologies formed after boiling water treatment entirely cover all surface features of micro scales originating out of the acid etching process. The presence of these highly conforming nano petals structure is highly desirable to generate hierarchical morphologies suitable for applications where adhesion is required to be minimized. Similarly, samples etched with 5 molar acid solutions show the presence of nano petal morphology when acid etching time was kept equal to or below 5 minutes. Interestingly samples which showed smoothening of nanoscale features due to higher acid concentration and higher etching time reveal very limited (see inset image 6.(j)) or complete absence of nano-petal morphology (see inset of figure 6.(k)). These observations could be explained by the absence of unreacted aluminium in cases with higher acid concentration and higher etching time, as unreacted aluminium is required to form such nano-petal morphology made of boehmite following the boiling water treatment.[25]

The reactions are as follows,

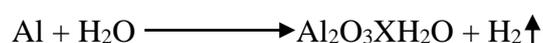

$$Al + H_2O \longrightarrow Al_2O_3XH_2O + H_2\uparrow$$

Based on the theoretical analysis and contact angle data of krytox, surfaces etched with 2.5 molar solution and an etching time of 20 minutes with/without subsequent boiling water treatment were selected for further study.

### 3.4 Surface energy modification

Selected texture surfaces were dip coated in liquid Teflon solution to modify the surface chemistry. As received, one weight percent Teflon solution was diluted to 0.1 weight percent by adding a fluorinated solvent FC72. Dip coating was performed with the diluted solution to ensure conformal coating, as high-concentration Teflon solution was found out to be non-conformal at sub-micron scale. Due to the higher thickness of coating obtained from high concentration Teflon solution, all submicron scale surface features were embedded within the coating. Hence the dilute solution of 0.1 weight percent was selected for dip coating. Dip coating with such low concentration may lead to local defects where the surface remains uncoated. Hence surfaces were coated in multiple passes, and surface wettability after each pass was measured to determine the number of passes corresponding to the maximum contact angle and minimum surface roll of angle. The table below shows the variation of these two parameters on Acid Etched Boiling Water (AEBW) treated samples for different number passes. As shown in table 1, samples with three passes showed low Roll-off Angle (RoA) and high ECA and hence were selected for dip coating of all selected surfaces. The ECA of water and droplet RoA on these three-pass coated was shown in figure 7.

Table 1: ECA and RoA of Al etched Sample (at Temp.25°C and RH 75)

| No of pass | ECA (°) | RoA (°) |
|---|---|---|
| 1 | 150 | 20 |
| 2 | 152 | 5 |
| 3 | 160 | 2.5 |
| 4 | 165 | 4 |
| 5 | 160 | 7 |

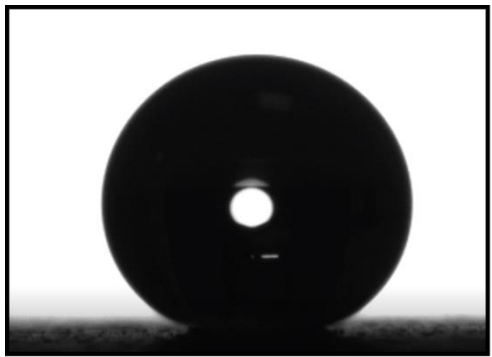
$\theta_E = 160°$

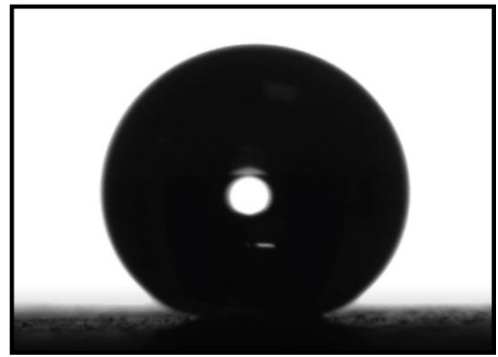
$\theta_{RoA} = 2.5°$

Figure 7: ECA and RoA of textured Teflon-coated three passes.

### 3.5 Stability of Lubricant

The experimental method described elsewhere (see experimental section) was used to observe the presence of a lubricant oil layer on the LIS surface submerged under different environmental fluids. As shown in the figure below oil phase remained on LIS surfaces even after one month inside cyclopentane. The shiny layer observed on the left side samples indicates the presence of an oil phase, whereas Teflon coated non-LIS surface with a similar surface texture appears much darker. Both surfaces with micro-only features and micro-nano hierarchical textures retained this lubricant oil layer for at least one month (maximum duration tested) under a cyclopentane environment. Similar observations were also made when these surfaces were kept inside water for the same duration. This observation not only validates the theoretical predication but also show potential superior performance of these surfaces in reducing hydrate adhesion.

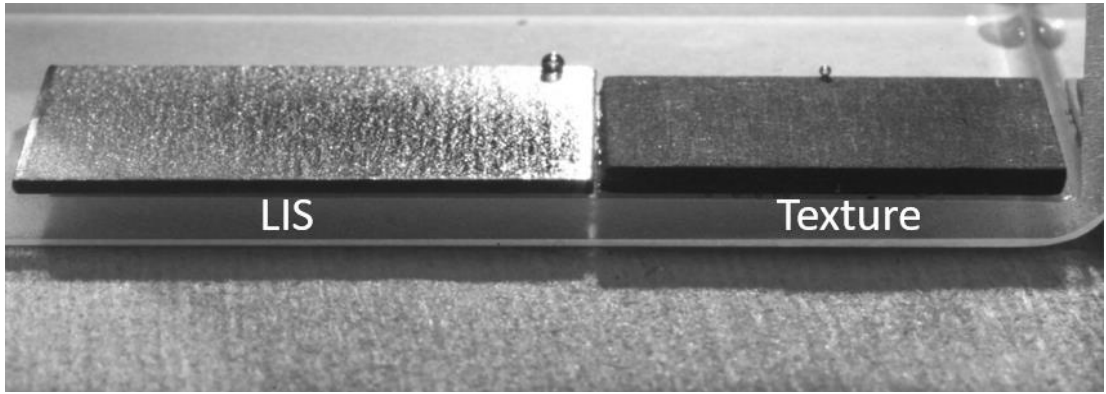

Figure 8: Stability of LIS surface

**3.6 Hydrate adhesion**

Hydrate adhesion measurements were performed on various test surfaces following the experimental protocol elsewhere. As expected, hydrate adhesion forces are reduced by more than one order of magnitude value compared to the smooth mirror-finished aluminium alloy. Figure 9. (a) and (b) shows the position of the cantilever just before and after the adhesion fracture of adjacent hydrate particle. The displacement between these two positions was practically zero indicating the adhesion force to be less than 0.041 mN. Right side of the below figure, shows the bar plot of hydrate adhesion forces on different test surfaces. The extremely low hydrate adhesion on LIS with micro-nano hierarchical structure can be explained due to the low hydrate to the solid contact area and the presence of lubricant barrier film in between them. Hydrate adhesion force seems unaffected by the increase of lubricant oil viscosity examined in this report.

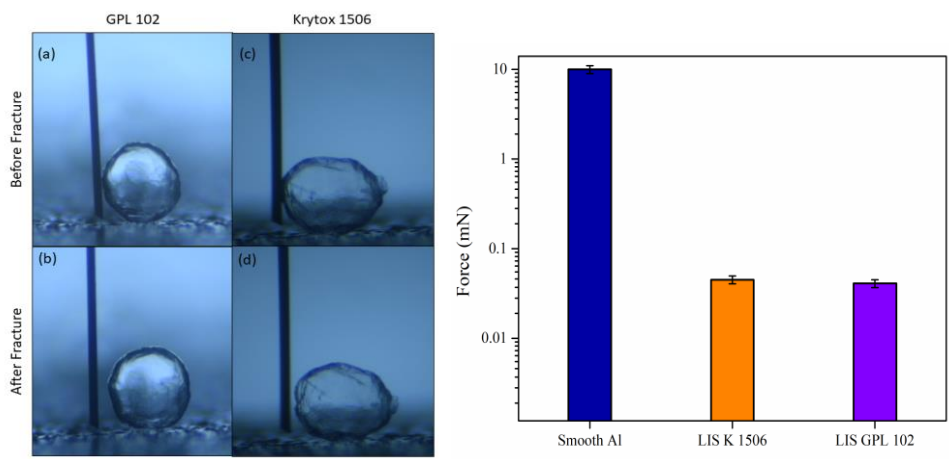

Figure 9: Steel cantilever wire used to dislodge the hydrate particles from the test surface on the left side and hydrate adhesion force on the right side.

## 4. Conclusions

In this current report, Liquid Impregnated Surfaces (LIS) based on industrial material and industrially scalable process with extremely low hydrate adhesion has been designed, fabricated, and experimentally demonstrated. Best performing LIS showed at least one order of magnitude less hydrate adhesion force compared to the lowest hydrate adhesion force reported till date in literature.[26] A systematic design approach was followed to select suitable surface chemistry and surface textured with unique nano petal features to fabricate LIS. Theoretically predicted stability of lubricant oil on these LIS was experimentally validated. Subsequent measurement of adhesion forces on these LIS demonstrates the accuracy of the design approach and robustness of these surfaces in reducing hydrate adhesion with extremely low value of 0.5 KN/m$^2$. Due to extremely low hydrate adhesion and the use of industrially scalable low-cost materials fabrication method, these surfaces have tremendous potential for real-life application.

Low hydrate adhesion on these LIS is a cumulative effect of interfacial energies between different phases. The surfactant, anti-agglomerate, and other chemicals used in industry for anticlogging may significantly alter the interfacial interaction between water oil lubricant and solid phase. The altered interfacial interaction is bound to affect the hydrate adhesion too. The performance of LIS surfaces in such scenarios is worth further investigation. Moreover, the suitability of the design approach and theoretical calculation reported in this article can be experimentally investigated in the modified systems. Lubricant oils used in LIS may also affect the cohesion force between hydrate particles, which needs further investigation. The lubricant layer may also act as a reservoir for other additives to reduce adhesion and cohesion forces in the hydrate system. Nevertheless, LIS possesses the unique advantage of having thermodynamically stable lubricant oil layers unlike other surfaces such as SLIPS and Superhydrophobic surfaces where lubricant oil layer or air layer stability is not always ensured. The stability of lubricant oil layers on LIS under flow conditions has also not been investigated. Future research work in these directions may give clear idea on feasibility of these surfaces to be used in oil and gas pipelines under flow condition.

It is necessary to carry out a stability study of this micro-nano roughness in order to retain the functional durability over a long period of time. According to the FE-SEM micrograph, the height, width, and spacing between the roughness are all in the range of 100 microns. When compared to an open pore structure, these dead pores additionally offer improved oil retention. Due to the higher aspect ratio pore structure, which significantly lowers the ECA of oil on solid in a water environment and enhances oil stability, the lubricating oil layer that is present on top of this micro-nano scale roughness is stable. However, disjoining pressure was not taken into account in the theoretical calculation. With this idea, overall thermodynamic stability may be ensured.

## 5 Materials and Methods

### 5.1 Materials

The mirror-finished aluminium alloy 6061 was purchased from Mac-Master USA. The liquid Teflon (polytetrafluoroethylene) was purchased from Sigma-Aldrich, and HCl (hydrochloric acid) was purchased from Merck India Pvt. Ltd. The FC-72 (tetradecaflurohexane) was purchased from DuPont. Acetone, Isopropanol (IPA) and ethanol were purchased from Merck India Pvt. Ltd. Deionised water is prepared by the Millipore water purification system.

### 5.2 Surface Texturing

Aluminium 6061, a multipurpose alloy of aluminium was selected for surface texturing. Untreated surfaces were of mirror finish. To ensure cleanliness, substrate ware bath sonicated in acetone, Isopropanol, and DI water. Different micro-structures on this aluminium samples were generated using the Hydrochloric acid-based chemical etching method. Subsequently, samples were treated in boiling water to generate nano textures. To evaluate the effect of acid concentration, etching processes were performed with 0.1M, 1M, 2.5M, and 5M HCl solutions. Samples from each of these cases were taken out from the etching bath at different times of 5, 10, 15, and 20 minutes to evaluate the influence of reaction time on microfeature generation. After collecting from the acid bath, samples were rinsed with copious amounts of water. Cleaned surfaces were subsequently kept inside boiling water for 20 minutes. After boiling water treatment, samples are dried in a convection oven and stored in a vacuum desiccator before further processing.

## 5.3 Surface Energy Modification

The surface energy of both smooth and texture aluminium surfaces were modified by dip coating in liquid Teflon solution using a precision dip coater. Teflon concentration, Dip coating speed, and number of coating passes were varied to optimum suitable values of these parameters for the synthesis of nanoscale conforming coatings. As received, Teflon solution was diluted using FC 72, a fluorinated volatile solvent dip-coated samples were air dried and kept on the hot plate for heat curing. The samples are first heat cured initially at 160°C for 5 minutes and followed by a second curing state at 240°C for 15 minutes. Samples which were coated in multiple passes were heat curried after each pass.

## 5.4 Surface characterization

### 5.4.1 Wettability

Both static and dynamic contact angles along with the roll of angles on different surfaces, were measured using a contact angle goniometer (Rame-Hart model 500-U1). To measure the contact angle of lubricant inside water and cyclopentane environment, an optically transparent quartz cell was used. Test samples kept inside this quartz cell were prefilled with environmental fluid (water or cyclopentane) for subsequent contact angle measurements of lubricant (krytox 1506, GPL 102).

### 5.4.2 Surface morphology

Surface morphologies of textured aluminium surfaces were characterize using a Field emission scanning electron microscope (FE-SEM, Sigma 300 Carl-Zeiss). FE-SEM images were taken in different magnifications to visualize surface feature at micro and nano-scale.

### 5.5 Liquid Impregnation of textured solid

Liquid impregnated surfaces were prepared by infusing lubricant oil on Teflon-coated textured aluminium surfaces using a dip coating method. A dip coating speed of 1 mm/min was selected to ensure the minimal thickness of the excess lubricant layer.

### 5.6 Lubricant Stability Experiment

To experimentally assess the stability of lubricant oil of LIS surfaces under the water and cyclopentane environment, a custom-made experimental setup was used. A transparent quartz cell of dimensions 7cm by 7cm by 7cm was filled with environmental liquid. Subsequently, the test LIS surface, Textured Teflon-coated surface, and textured aluminium surface were placed inside the liquid-filled quartz cell. A DSLR camera and a suitable light source were appropriately placed to visualize the lubricant or plastron layer. Images were taken until one month to detect any instability of the lubricant layer (plastron layer).

### 5.7 Hydrate adhesion measurement

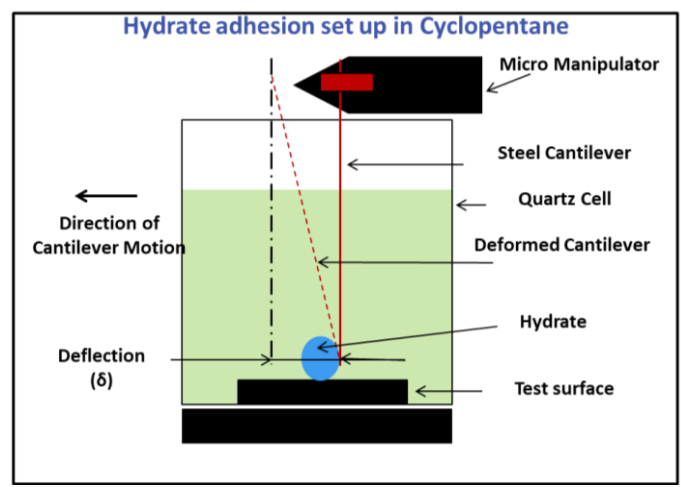

Figure 10: Schematic diagram of the hydrate measurement setup.

Adhesion between hydrate particles and test surfaces was measured using a custom-made experimental setup comprising of computer controlled Peltier module, optically transparent glass cuvette, DSLR camera, light source, and thin cantilever wire mounted on a micro-manipulator system. The micro-manipulator and cantilever system was attached with a highly accurate custom-made XYZ linear stage for proper alignment and positioning.

In this experimental method, hydrate particles were formed at the interface between the tiny water drop and cyclopentane environment. A glass cuvette containing a test sample submerged under cyclopentane was kept on the Peltier cooler. A few millimeter scale water drops were deposited in a side glass cuvette (not on test surfaces) for hydrate seeding purposes to make cyclopentane hydrate. Subsequently, the temperature of the cyclopentane was reduced to $-15°C$ using the Peltier cooler. Once the deposited water drop was frozen, the temperature of the Peltier cooler was increased to $1°C$ with a very slow rate in such a way that the ice melting rate is very low. This low melting rate of ice ensures rapid nucleation of cyclopentane hydrates at the water cyclopentane interface. The formation of the hydrate phase was visually detectable in a DSLR camera. Once the hydrate phase entirely covered the water cyclopentane interface, the temperature of the system was raised to $3°C$ to ensure no ice phase remained. Tiny water drops of 0.5mm in diameter were deposited on the different locations of test surfaces. Subsequently, hydrate crystals were collected from the millimeter scale hydrate particles (formed on millimeter scale diameter) and transferred to tiny water droplets kept on test surfaces. The hydrate phase started growing from these crystals, and eventually, all tiny water droplets were converted into hydrate particles. The hydrate adhesion measurements between these hydrate particles and test surfaces were carried out using a thin steel cantilever of diameter 0.1mm after three hours of waiting time. The cantilever tip was properly aligned and positioned next to the individual hydrate particles in such a way that direction of cantilever motion during the measurement was always perpendicular to the optical axis. A DSLR camera (Nikon D 850) was used to record the events starting from the movement of cantilever touched hydrate particles till hydrate particles were completely detached from the surface or cohesively broken. The distance traveled by the cantilever beam during these two events was measured using Photoshop. This distance represents the deflection of the cantilever beam to be used for adhesion force measurements using the classical beam deflection method.

**Corresponding Author**

*Email: arindam@iitgoa.ac.in

**ORCID ID**

Amit Nayse: 0000-0002-2454-7604

Abhishek Mund: 0000-0003-2924-1804

Arindam Das: 0000-0002-8163-0666

**Author Contributions**

The manuscript was written through the contributions of all authors. All authors have given approval to the final version of the manuscript.

**Notes**

The authors declare no competing financial interest.

**References**

(1) Liu, W.; Hu, J.; Wu, K.; Sun, F.; Sun, Z.; Chu, H.; Li, X. A New Hydrate Deposition Prediction Model Considering Hydrate Shedding and Decomposition in Horizontal Gas-Dominated Pipelines. *Pet. Sci. Technol.* **2019**, *37* (12), 1370–1386. https://doi.org/10.1080/10916466.2019.1587457.

(2) Jr, E. D. S.; Koh, C. A.; Koh, C. A. *Clathrate Hydrates of Natural Gases*, 3rd ed.; CRC Press: Boca Raton, 2007. https://doi.org/10.1201/9781420008494.

(3) Castellani, B.; Morini, E.; Filipponi, M.; Nicolini, A.; Palombo, M.; Cotana, F.; Rossi, F. Clathrate Hydrates for Thermal Energy Storage in Buildings: Overview of Proper Hydrate-Forming Compounds. *Sustainability* **2014**, *6* (10), 6815–6829. https://doi.org/10.3390/su6106815.

(4) Sum, A. K.; Koh, C. A.; Sloan, E. D. Clathrate Hydrates: From Laboratory Science to Engineering Practice. *Ind. Eng. Chem. Res.* **2009**, *48* (16), 7457–7465. https://doi.org/10.1021/ie900679m.

(5) Aman, Z. M.; Leith, W. J.; Grasso, G. A.; Sloan, E. D.; Sum, A. K.; Koh, C. A. Adhesion Force between Cyclopentane Hydrate and Mineral Surfaces. *Langmuir* **2013**, *29* (50), 15551–15557. https://doi.org/10.1021/la403489q.

(6) Aman, Z. M.; Sloan, E. D.; Sum, A. K.; Koh, C. A. Adhesion Force Interactions between Cyclopentane Hydrate and Physically and Chemically Modified Surfaces. *Phys. Chem. Chem. Phys.* **2014**, *16* (45), 25121–25128. https://doi.org/10.1039/C4CP02927E.

(7) Wang, F.; Ma, R.; Xiao, S.; English, N. J.; He, J.; Zhang, Z. Anti-Gas Hydrate Surfaces: Perspectives, Progress and Prospects. *J. Mater. Chem. A* **2022**, *10* (2), 379–406. https://doi.org/10.1039/D1TA08965J.

(8) Nicholas, J. W.; Dieker, L. E.; Sloan, E. D.; Koh, C. A. Assessing the Feasibility of Hydrate Deposition on Pipeline Walls—Adhesion Force Measurements of Clathrate Hydrate Particles on Carbon Steel. *J. Colloid Interface Sci.* **2009**, *331* (2), 322–328. https://doi.org/10.1016/j.jcis.2008.11.070.

(9) Subramanyam, S. B.; Rykaczewski, K.; Varanasi, K. K. Ice Adhesion on Lubricant-Impregnated Textured Surfaces. *Langmuir* **2013**, *29* (44), 13414–13418. https://doi.org/10.1021/la402456c.

(10) Song, F.; Wu, C.; Chen, H.; Liu, Q.; Liu, J.; Chen, R.; Li, R.; Wang, J. Water-Repellent and Corrosion-Resistance Properties of Superhydrophobic and Lubricant-Infused Super Slippery Surfaces. *RSC Adv.* **2017**, *7* (70), 44239–44246. https://doi.org/10.1039/C7RA04816E.

(11) Solomon, B. R.; Khalil, K. S.; Varanasi, K. K. Drag Reduction Using Lubricant-Impregnated Surfaces in Viscous Laminar Flow. *Langmuir* **2014**, *30* (36), 10970–10976. https://doi.org/10.1021/la5021143.

(12) Schenkel, M. L.; Irazuzta, V. L.; Candia, G. A.; Scianca, N.; Aguero, S. D. Production Optimization: Understanding and Solving Scale Formation in Mature Gas Wells; OnePetro, 2020. https://doi.org/10.2118/201301-MS.

(13) Seo, D.; Shim, J.; Shin, D. H.; Nam, Y.; Lee, J. Dropwise Condensation of Acetone and Ethanol for a High-Performance Lubricant-Impregnated Thermosyphon. *Int. J. Heat Mass Transf.* **2021**, *181*, 121871. https://doi.org/10.1016/j.ijheatmasstransfer.2021.121871.

(14) Dong, S.; Li, M.; Liu, C.; Zhang, J.; Chen, G. Bio-Inspired Superhydrophobic Coating with Low Hydrate Adhesion for Hydrate Mitigation. *J. Bionic Eng.* **2020**, *17* (5), 1019–1028. https://doi.org/10.1007/s42235-020-0085-5.

(15) Li, S.; Lv, R.; Yan, Z.; Huang, F.; Zhang, X.; Chen, G.-J.; Yue, T. Design of Alanine-Rich Short Peptides as a Green Alternative of Gas Hydrate Inhibitors: Dual Methyl Group Docking for Efficient Adsorption on the Surface of Gas Hydrates. *ACS Sustain. Chem. Eng.* **2020**, *8* (10), 4256–4266. https://doi.org/10.1021/acssuschemeng.9b07701.

(16) Li, X.-M.; Reinhoudt, D.; Crego-Calama, M. What Do We Need for a Superhydrophobic Surface? A Review on the Recent Progress in the Preparation of Superhydrophobic Surfaces. *Chem. Soc. Rev.* **2007**, *36* (8), 1350–1368. https://doi.org/10.1039/B602486F.

(17) Quan, Y.-Y.; Chen, Z.; Lai, Y.; Huang, Z.-S.; Li, H. Recent Advances in Fabricating Durable Superhydrophobic Surfaces: A Review in the Aspects of Structures and Materials. *Mater. Chem. Front.* **2021**, *5* (4), 1655–1682. https://doi.org/10.1039/D0QM00485E.

(18) Smith, J. D.; Dhiman, R.; Anand, S.; Reza-Garduno, E.; Cohen, R. E.; McKinley, G. H.; Varanasi, K. K. Droplet Mobility on Lubricant-Impregnated Surfaces. *Soft Matter* **2013**, *9* (6), 1772–1780. https://doi.org/10.1039/C2SM27032C.

(19) Fan, S.; Zhang, H.; Yang, G.; Wang, Y.; Li, G.; Lang, X. Reduction Clathrate Hydrates Growth Rates and Adhesion Forces on Surfaces of Inorganic or Polymer Coatings. *Energy Fuels* **2020**, *34* (11), 13566–13579. https://doi.org/10.1021/acs.energyfuels.0c01904.

(20) Mund, A.; Nayse, A. K.; Das, A. Design of a Liquid Impregnated Surface with a Stable Lubricant Layer in a Mixed Water/Oil Environment for Low Hydrate Adhesion. *Langmuir* **2023**, *39* (34), 11964–11974. https://doi.org/10.1021/acs.langmuir.3c00320.

(21) Karanjkar, P. U.; Lee, J. W.; Morris, J. F. Surfactant Effects on Hydrate Crystallization at the Water–Oil Interface: Hollow-Conical Crystals. *Cryst. Growth Des.* **2012**, *12* (8), 3817–3824. https://doi.org/10.1021/cg300255g.

(22) Wang, F.; Xiao, S.; He, J.; Ning, F.; Ma, R.; He, J.; Zhang, Z. Onion Inspired Hydrate-Phobic Surfaces. *Chem. Eng. J.* **2022**, *437*, 135274. https://doi.org/10.1016/j.cej.2022.135274.

(23) Brown, E.; Hu, S.; Wang, S.; Wells, J.; Nakatsuka, M.; Veedu, V.; Koh, C. Low-Adhesion Coatings as a Novel Gas Hydrate Mitigation Strategy; OnePetro, 2017. https://doi.org/10.4043/27874-MS.

(24) Morrissy, S. A.; Lim, V. W.; May, E. F.; Johns, M. L.; Aman, Z. M.; Graham, B. F. Micromechanical Cohesive Force Measurements between Precipitated Asphaltene

Solids and Cyclopentane Hydrates. *Energy Fuels* **2015**, *29* (10), 6277–6285. https://doi.org/10.1021/acs.energyfuels.5b01427.

(25) Aizawa, T.; Yoshino, T.; Inohara, T. Micro-/Nano-Texturing of Aluminum by Precise Coining for Functional Surface Decoration. *Metals* **2020**, *10* (8), 1044. https://doi.org/10.3390/met10081044.

(26) Das, A.; Farnham, T. A.; Bengaluru Subramanyam, S.; Varanasi, K. K. Designing Ultra-Low Hydrate Adhesion Surfaces by Interfacial Spreading of Water-Immiscible Barrier Films. *ACS Appl. Mater. Interfaces* **2017**, *9* (25), 21496–21502. https://doi.org/10.1021/acsami.7b00223.